\documentclass{sig-alternate-10pt}
\usepackage[T1]{fontenc}
\usepackage{lmodern}
\usepackage[utf8]{inputenc}
\PassOptionsToPackage{hyphens}{url}
\usepackage[unicode]{hyperref}
\usepackage{xspace}
\usepackage{caption}
\usepackage{color}
\usepackage{graphicx}
\usepackage{booktabs}
\usepackage{multirow}
\usepackage{tumcolor}
\usepackage{tabularx}
\usepackage[group-digits=true, group-separator={,}]{siunitx}
\usepackage[pointedenum]{paralist}
\usepackage{url}
\PassOptionsToPackage{hyphens}{url}
\usepackage{listings}
\usepackage{cleveref}
\usepackage{comment}
\usepackage{subcaption} 
\usepackage{subfloat} 
\usepackage{pifont} 
\usepackage{fancyhdr}
\usepackage{algorithmic}
\usepackage{algorithm}

\algsetup{linenodelimiter=}

\DeclareSIUnit{\nothing}{\relax}

\graphicspath{{./}{./figures/}{./RFigures/}}

\usepackage{microtype}
\microtypesetup{expansion=alltext,step=1} 
\newif\ifstatus
\statustrue
\hypersetup{pdfstartview=FitV,
  colorlinks=true, linkcolor=black, citecolor=black, urlcolor=blue,
}

\newcounter{nalg}[section] 
\renewcommand{\thenalg}{\thesection .\arabic{nalg}} 
\DeclareCaptionLabelFormat{algocaption}{Algorithm \thenalg} 

\newcommand{\xref}[1]{\hyperref[#1]{\S\ref*{#1}}\xspace}

\begin{document}
\pagestyle{fancy}
\lhead{}
\cfoot{\textcolor{TUMDarkerBlue}{Do you archive a top list and want to help future studies by sharing this with us? Then please contact us!}}
\rfoot{\thepage}
\chead{}
\rhead{}
\renewcommand{\headrulewidth}{0.0pt}

\title{Structure and Stability of Internet Top Lists}
\vspace{-10mm}

\numberofauthors{6}
\author{
\alignauthor
Quirin Scheitle\\
       \affaddr{Technical University of Munich}\\
\alignauthor
Jonas Jelten\\
       \affaddr{Technical University of Munich}\\
\alignauthor
Oliver Hohlfeld\\
       \affaddr{RWTH Aachen University}\\
\and 
\alignauthor
Luca Ciprian\\
       \affaddr{Technical University of Munich}\\
\alignauthor
Georg Carle\\
       \affaddr{Technical University of Munich}\\
}
\date{\today}

\maketitle
\begin{abstract}%
Active Internet measurement studies rely on a list of targets to be scanned. 
While probing the entire IPv4 address space is feasible for scans of limited complexity, more complex scans do not scale to measuring the full Internet.
Thus, a sample of the Internet can be used instead, often in form of a ``top list''. 
The most widely used list is the Alexa Global Top1M list. 
Despite their prevalence, use of top lists is seldomly questioned.
Little is known about their creation, representativity, potential biases, stability, or overlap between lists.
As a result, potential consequences of applying top lists in research are not known.
In this study, we aim to open the discussion on top lists by investigating the aptness of frequently used top lists for empirical Internet scans, including stability, correlation, and potential biases of such lists.
\end{abstract}\vspace{-2mm}
\section{Lists of Popular Domains}\label{sec:intro}
Internet Top lists contain frequently accessed domains according to typically proprietary data by the list creator.
The following lists are widely used: 

\textbf{Alexa} top lists~\cite{alexa} are created based on usage data collected by the Alexa browser plugin.
No information exists on the plugin's user base and thus opens questions on list representativity and potential biases (towards the plugins' unknown user base).
Alexa lists are offered for sale with few free offerings.
The most popular free offering is the list of the global top 1M domains.
Paid offerings include top lists per country, industry, or region.
For each, the top 50 entries can be viewed free of charge.

\textbf{Quantcast}~\cite{quantcast} provides a list of the top 1\,M most frequently visited web sites per country, measured through their web intelligence plugin on sites.
Only the US-based list can be downloaded, all other lists can only be viewed online and hide ranks when not purchased.
Thus, we do not systematically download and evaluate it.

\textbf{Majestic Million}~\cite{majestic} offers a creative commons licensed top 1\,M list based on Majestic's web crawler, which ranks sites by the number of subnets linking to that site. 
This is a different data collection methodology, and similar to Alexa, heavily web-focused.

\textbf{Cisco Umbrella}~\cite{umbrella} contains the list of top 1\,M domains (including sub domains) according to DNS queries by users of Cisco's OpenDNS system.
This is a fundamentally different nature than collecting web site visits or links, as it is based on DNS requests for all kinds of Internet services, not just web sites.

\textbf{Top List Use in Research.}
We start by studying the use of top lists in measurement research among 3 Internet Measurement conferences (i.e., IMC, PAM, and TMA) in 2015--2017.
Out of 260 papers published at these conferences, 56 (21.5\%) utilize a top list (see Table~\ref{tab:toplistusage}).
We find {\em all} 56 papers to use an Alexa list, while two papers additionally use either the Cisco Umbrella or the Quantcast list. 
Of these 56 papers, 48 use the global list, 8 a country-specific list, and 6 categorical lists, with some papers using several of these.
Two papers only state to use ``the Alexa list''.

\renewcommand*{\thesubfigure}{1\alph{subfigure}}
\begin{figure*}[t]
	\begin{subfigure}[t]{0.33\textwidth}
		\includegraphics[width=\columnwidth]{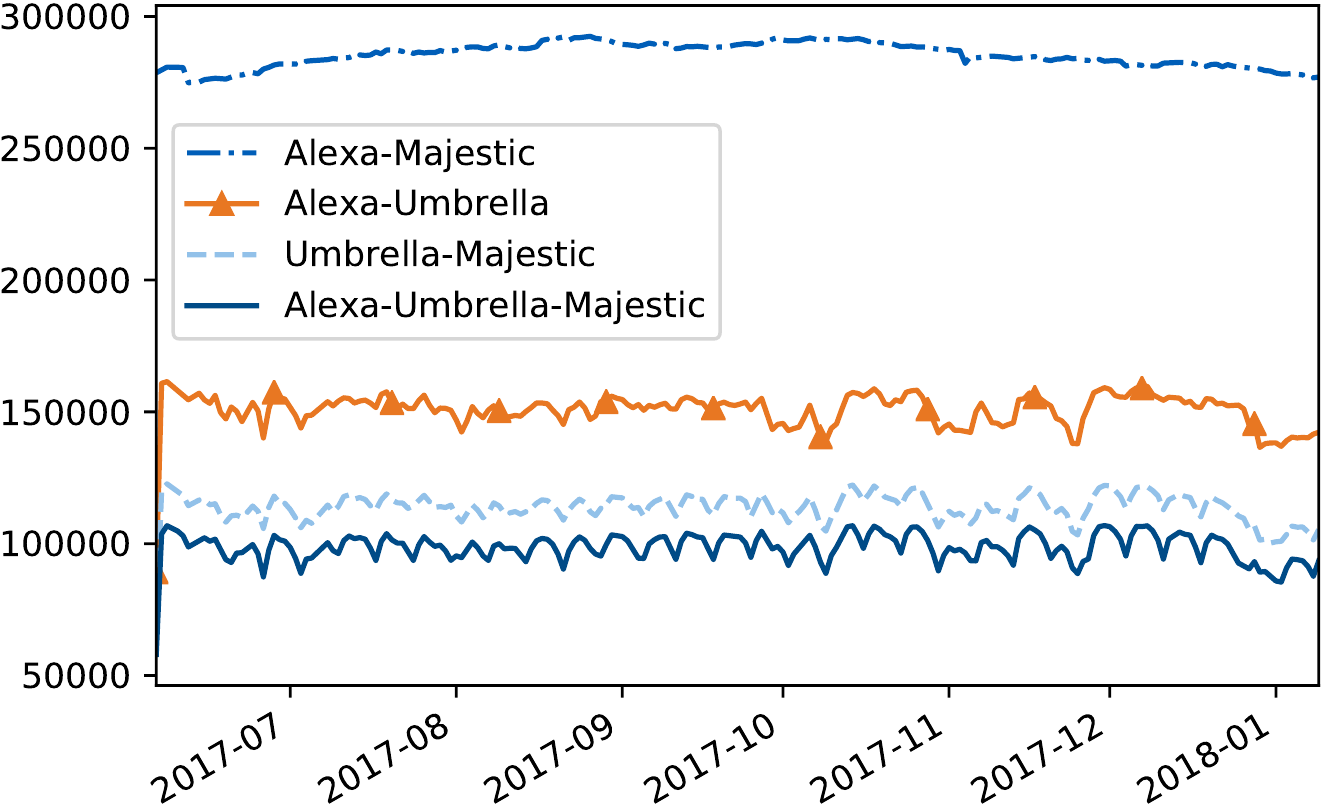}
		\vspace{-5mm}
		\caption{Intersection between top lists}
		\label{DISABLEDsubfig:intersection}
	\end{subfigure}
	\hfill
	\begin{subfigure}[t]{0.32\textwidth}
		\includegraphics[width=\columnwidth]{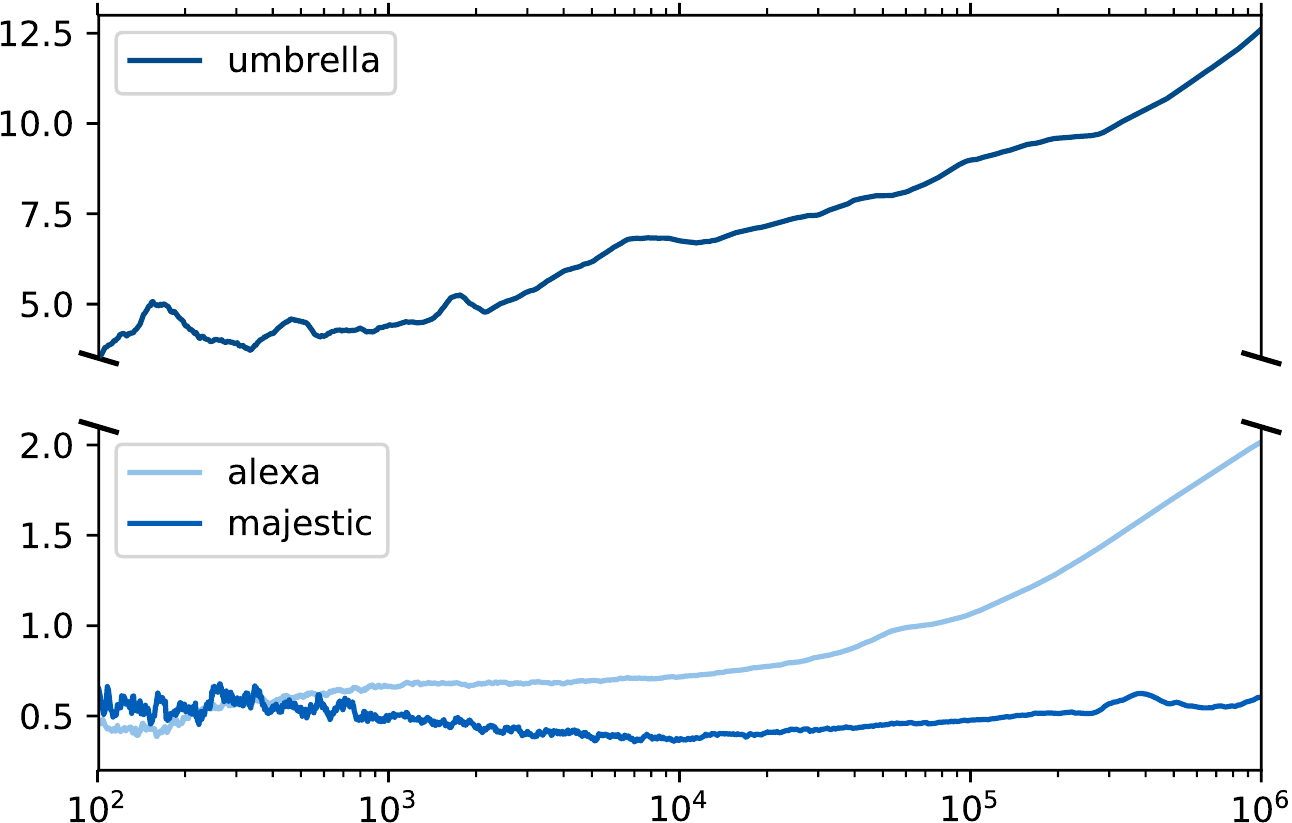}
		\vspace{-5mm}       
		\caption{Average \% Daily Change over Rank}
		\label{DISABLEDsubfig:dailychangerank}
	\end{subfigure}
	\hfill
	\begin{subfigure}[t]{0.3\textwidth}
		\includegraphics[width=\columnwidth, height=0.7\columnwidth, keepaspectratio]{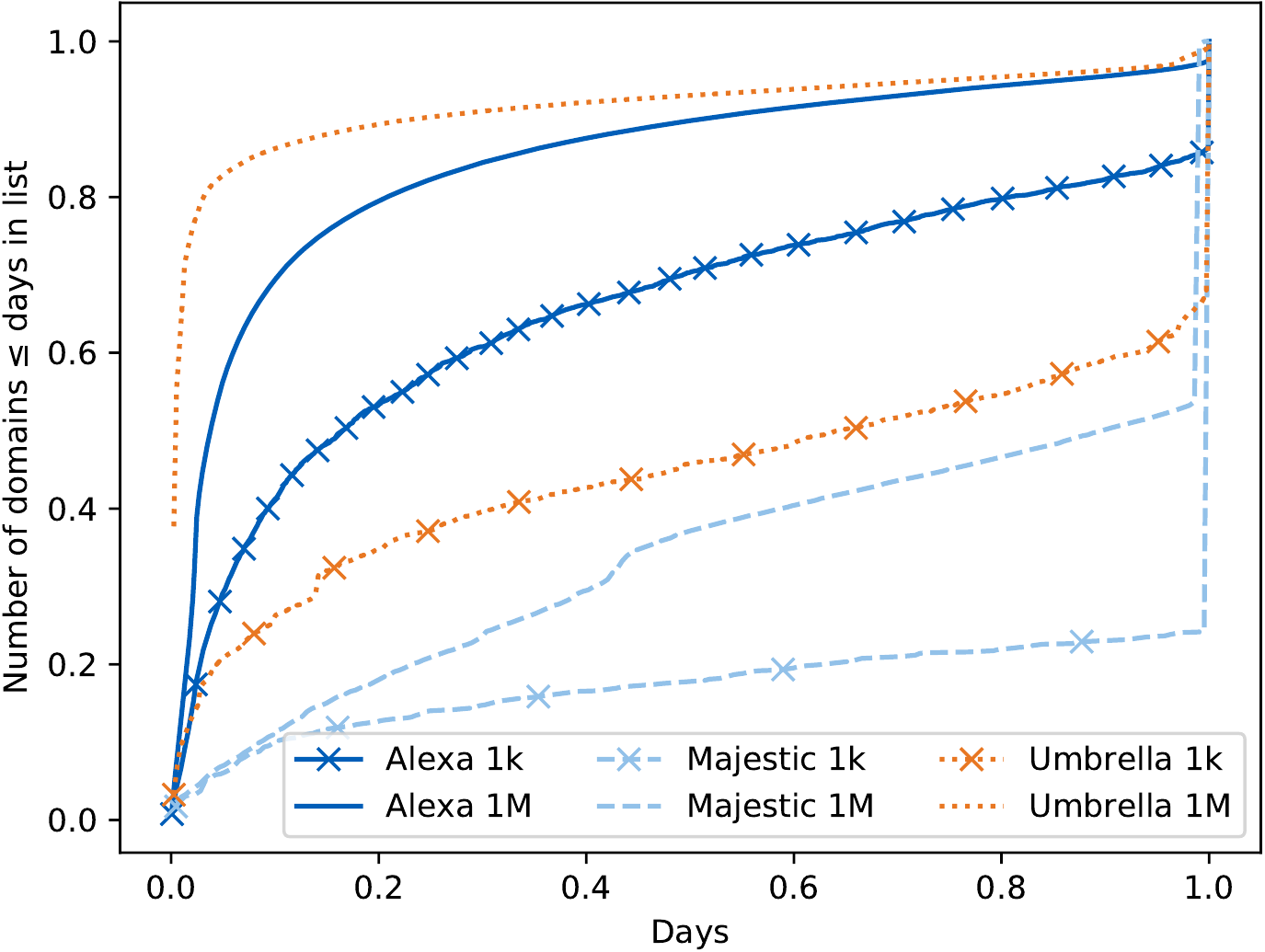}
		\vspace{-1mm}
		\caption{CDF of domains over days in top lists}
		\label{DISABLEDsubfig:ndays}
	\end{subfigure}
\end{figure*}

\begin{table}
\small
\setlength{\tabcolsep}{3pt}
\begin{tabular}{lrllrrlr}
\toprule
\multirow{2}{*}{List} & \multicolumn{3}{c}{\# Papers}                                             & \multicolumn{4}{c}{Alexa Only}                                   \\
                      & \multicolumn{1}{l}{IMC} & PAM                    & TMA                    & \multicolumn{2}{c}{Global Top} & \multicolumn{2}{c}{Specific}    \\
\cmidrule{1-4}\cmidrule(l){5-8}
Alexa~\cite{alexa}                & 35                      & \multicolumn{1}{r}{13} & \multicolumn{1}{r}{8}  & 500             & 7            & Country  & 8                    \\
Umbrella~\cite{umbrella}              & 1                       & \multicolumn{1}{r}{--} & \multicolumn{1}{r}{--} & 1k              & 5            & Category & 6                    \\
Quantcast~\cite{quantcast}             & 1                       & \multicolumn{1}{r}{--} & \multicolumn{1}{r}{--} & 10k             & 3            & Global   & 48                   \\
                      &                         &                        &                        & 100k            & 4            &          & \multicolumn{1}{l}{} \\
                      & \multicolumn{1}{l}{}    &                        &                        & 1M              & 27           &          & \multicolumn{1}{l}{} \\
\bottomrule
\end{tabular}
\vspace{-3mm}
\caption{\small Use of top lists in 260 papers published at IMC, PAM, and TMA in 2015, 2016, and 2017.}
\label{tab:toplistusage}
\vspace{-6mm}
\end{table}
\vspace{-3mm}\section{Structure: Subdomain Depth}\vspace{-1mm}
Top lists vary in the provided level of detail in terms of subdomain depth.
For example, for \textit{www.net.in.tum.de}, \textit{.de} is the public suffix, \textit{tum.de} is the base domain, 
\textit{in.tum.de} is the first subdomain, and \textit{net.in.tum.de} is the second subdomain.
We count the list entry \textit{www.net.in.tum.de} as a third-level subdomain.
\Cref{tab:data} shows the average number of base domains ($\mu_{BD}$) per top list. 
We note that Alexa and Majestic contain almost exclusively base domains with few exceptions (e.g., for blogspot).
 In contrast, Umbrella only contains an average of 28\% base domains.
\Cref{tab:subdomains} details the subdomain depth for a single-day snapshot of all lists.
Umbrella holds deep subdomains levels, up to level 33 (an IPv6 rDNS pointer).
We also note that the base domain is usually part of the list when its subdomains are listed:
On average each list contains only few hundred subdomains whose base domain is not part of the list. 
\newline Thus, the choice of lists can be based on the desired depth: Choose Umbrella when subdomains are needed.

\begin{table}
	\centering
	\resizebox{\columnwidth}{!}
	{
		\begin{tabular}{lrrr}
			\toprule
			List & Since & $\mu_{BD}\pm\sigma$& $\mu_\Delta\pm\sigma$\\
			\midrule
			Alexa Top1M~\cite{alexa} & 9.6.14 & 976k $\pm$ 908 & 20k $\pm$12k\\
			Cisco Umbrella~\cite{umbrella}& 15.12.16 & 278k $\pm$ 14k & 126k $\pm$118k\\
			Majestic Million~\cite{majestic}& 6.6.17 & 994k $\pm$ 652 & 6k $\pm$4k\\
			\bottomrule
		\end{tabular}
	}
	\vspace{-3mm}
	\caption{\small Top Lists Datasets, mean of base domains ($\mu_{BD}$) and mean of daily change ($\mu_\Delta$) of raw domains.}
	\label{tab:data}
	\vspace{-3mm}
\end{table}

\section{Low Intersection between lists}
We next study intersection between lists---small intersections suggest a bias in list creations.
Figure 1a shows the intersection between top lists over time.
We see that the intersection is quite small:
Alexa and Majestic share 285k domains on average, Alexa and Umbrella 150k, Umbrella and Majestic 113k, and all three only agree on 99k out of 1M domains.
This disagreement on top domains suggests a high bias in the list creation.
We note that both web-based lists, Alexa and Majestic, only share an average of 29\% of domains.
Thus, Alexa provides {\em one} answer to which domains are popular, while other lists provide quite a {\em different} answer.

\begin{table}
	\centering
	\small
	\begin{tabular}{lrrr}
		\toprule
		 Domain Level & Alexa & Umbrella & Majestic \\
		\midrule
		\% Base Domains & 97.6\% & 24.9\% & 99.5\% \\
		\% 1st subdomain & 2.35\% & 54.4\% & 0.4\% \\
		\% 2nd subdomain &  0.01 \% & 13.3\% &  $\approx$0 \\
		\% 3rd subdomain & $\approx$0 & 5.5\% &$\approx$0 \\
		Max subdomain level & 4 & 33 & 4 \\
		\bottomrule
	\end{tabular}
	\vspace{-3mm}
	\caption{\small Subdomain depth as of Jan-9-2018.}
	\vspace{-5mm}
	\label{tab:subdomains}
\end{table}
\vspace{-2mm}\section{Stability Varies with List \& Rank}
An important aspect is list stability: Were the results of studies significantly different if they had picked a list from a different date? 
We first look at the number of daily changes ($\mu_\Delta$) per full list in \Cref{tab:data}. 
We find Umbrella quite unstable at 12.6\% average churn per day, but Alexa and Majestic quite stable at 0.6 to 2\%.

In Figure 1b we investigate the stability of lists in dependence of ranks. 
We compute the average daily churn of domain subsets increasing by rank for each list.
The figure shows instability increasing with higher ranks for Alexa and Umbrella, but not for Majestic. 
This confirms the belief that the Alexa bottom ranks are rather unstable, though at a moderate churn rate of 2\%.

We also investigate the average number of days a domain remains in a top list in Figure 1c. 
This figure displays a CDF with normalized days in the x-axis, and the normalized cumulative probability that a domain is X or less days part of the list. 
The lists show quite different behavior, with Majestic Top1K being the most stable by far ($\approx20\%$  domains present $\leq100\%$ days), and being followed by Majestic Top1M, Umbrella 1K, Alexa 1K, Alexa 1M and Umbrella 1M. 
Please also note that the Majestic 1M list is more stable than the Alexa and Umbrella Top1K lists.
The data sets cover differently long time periods (hence the normalization), however we consider even the shortest length of about 6 months as long enough to allow for reliable analysis.

\vspace{-2mm}\section{Insights and Further Questions}
The dominant use of the Alexa top lists aids comparability among studies and helps focusing on frequently visited and hence maybe more important parts of the Internet. 
However, we find the intersection between the three top lists surprisingly low, at a maximum of about 30\% for Alexa and Majestic, suggesting that there exists no uniform belief in what the top1M domains in the Internet would be.
We also find stability of Majestic Million astounding, at 0.6\% daily change rate. Alexa is slightly worse at 2\%, but Umbrella is quite unstable with 12.6\% average daily churn rate. We also find the deep subdomain level of Umbrella surprising. 
We see these insights as a start for a more extensive study of these top lists and their influence on Internet measurement studies. 
A main aspect for further analysis will be structural diversity: How many domains are located at the same hosting providers or CDNs? How many are aliases or redirections to the same page (e.g., same base domains with various ccTLDs)?

\pagestyle{plain}
\textbf{Peculiar Findings}: Some domains stood out in our analysis, for example the Alexa list contains full URLs such as \textit{youtube.com/users/abcdef}. 
In the Umbrella list, we found several IPv6 rDNS addresses.
{\small
\bibliographystyle{plain}
\vspace{-3mm}
\bibliography{paper}
\vspace{1mm}
}
\end{document}